\documentclass[aps, prl, reprint, superscriptaddress, twocolumnt]{revtex4-1}
\usepackage{amsmath} 
\usepackage{graphicx} 
\usepackage[OT4]{fontenc}
\usepackage{color}
\usepackage{bm}
\usepackage{upgreek}
\usepackage{braket}
\makeindex
\begin{document}
\title{Engineering Long-Lived Collective Dark States in Spin Ensembles} 
\author{Stefan Putz}
\email{stef.putz@gmail.com}
\affiliation{Vienna Center for Quantum Science and Technology, Atominstitut, TU Wien, Stadionallee 2, 1020 Vienna, Austria}
\affiliation{Zentrum f\"ur Mikro- und Nanostrukturen, TU Wien, Floragasse 7, 1040 Vienna, Austria}
\author{Andreas Angerer}
\affiliation{Vienna Center for Quantum Science and Technology, Atominstitut, TU Wien, Stadionallee 2, 1020 Vienna, Austria}
\affiliation{Zentrum f\"ur Mikro- und Nanostrukturen, TU Wien, Floragasse 7, 1040 Vienna, Austria}
\author{Dmitry O. Krimer}
\affiliation{Institute for Theoretical Physics, TU Wien, Wiedner Hauptstra{\ss}e 8-10/136, 1040 Vienna, Austria}
\author{Ralph Glattauer}   
\affiliation{Vienna Center for Quantum Science and Technology, Atominstitut, TU Wien, Stadionallee 2, 1020 Vienna, Austria}
\author{William J. Munro}   
\affiliation{NTT Basic Research Laboratories, 3-1 Morinosato-Wakamiya, Atsugi, Kanagawa 243-0198, Japan}
\author{Stefan Rotter}   
\affiliation{Institute for Theoretical Physics, TU Wien, Wiedner Hauptstra{\ss}e 8-10/136, 1040 Vienna, Austria}
\author{J\"org Schmiedmayer}
\affiliation{Vienna Center for Quantum Science and Technology, Atominstitut, TU Wien, Stadionallee 2, 1020 Vienna, Austria}
\affiliation{Zentrum f\"ur Mikro- und Nanostrukturen, TU Wien, Floragasse 7, 1040 Vienna, Austria}
\author{Johannes Majer}
\affiliation{Vienna Center for Quantum Science and Technology, Atominstitut, TU Wien, Stadionallee 2, 1020 Vienna, Austria}
\affiliation{Zentrum f\"ur Mikro- und Nanostrukturen, TU Wien, Floragasse 7, 1040 Vienna, Austria}
\date{\today}
\label{par:abstract}
\begin{abstract}
\linespread{1.1}
Ensembles of electron spins in hybrid microwave systems \cite{xiang_hybrid_2013} are powerful and versatile components for future quantum technologies. Quantum memories with high storage capacities \cite{nunn_multimode_2008} are one such example which require long-lived states that can be addressed and manipulated coherently within the inhomogeneously broadened ensemble. This broadening is essential for true multimode memories, but induces a considerable spin dephasing and together with dissipation from a cavity interface \cite{kubo_strong_2010,amsuss_cavity_2011,putz_protecting_2014} poses a constraint on the memory's storage time. In this work we show how to overcome both of these limitations through the engineering of long-lived dark states in an ensemble of electron spins hosted by nitrogen-vacancy centres in diamond.~By burning narrow spectral holes into a spin ensemble strongly coupled to a superconducting microwave cavity, we observe long-lived Rabi oscillations with high visibility and a decay rate that is a factor of forty smaller than the spin ensemble linewidth and thereby a factor of more than three below the pure cavity dissipation rate. This significant reduction lives up to the promise of hybrid devices to perform better than their individual subcomponents. To demonstrate the potential of our approach we realise the first step towards a solid-state microwave spin multiplexer by engineering multiple long-lived dark states. Our results show that we can fully access the ``decoherence free" \cite{beige_driving_2000} subspace in our experiment and selectively prepare protected states \cite{plankensteiner_selective_2015} by spectral hole burning. This technique opens up the way for truly long-lived quantum memories, solid-state microwave frequency combs \cite{de_riedmatten_solid-state_2008}, optical to microwave quantum transducers \cite{stannigel_optomechanical_2010,blum_interfacing_2015} and spin squeezed states. Our approach also paves the way for a new class of cavity QED experiments with dense spin ensembles, where dipole spin-spin interactions become important and many-body phenomena \cite{hsieh_observation_2009,ma_uncovering_2014} will be directly accessible~on~achip.
\end{abstract}
\maketitle
A major breakthrough for engineered solid-state quantum systems was the first demonstration of coherent exchange of single energy quanta in a superconducting circuit \cite{nakamura_coherent_1999}. Since then these circuits have shown remarkable progress \cite{wallraff_strong_2004} with many quantum operations and applications having been demonstrated \cite{hofheinz_synthesizing_2009}. However a fundamental issue have been the short coherence times of these solid-state devices, in particular as compared to atomic systems \cite{langer_long-lived_2005}. A promising way to overcome this limitation is to use electron spins in semiconductor crystals which have shown remarkably long coherence times (up to almost one hour \cite{saeedi_room-temperature_2013}). These artificial atoms \cite{hanson_coherent_2008} are naturally long-lived quantum memories, yet easy manipulation has been challenging. This is where a suitable combination of different components to form ``hybrid" quantum systems has become a key strategy \cite{imamovgllu_cavity_2009}. The hybridisation of superconducting circuits with electron spin ensembles \cite{kubo_strong_2010,schuster_high-cooperativity_2010,amsuss_cavity_2011,probst_anisotropic_2013,zollitsch_high_2015,tabuchi_coherent_2015} has the potential to bypass the weaknesses of the individual systems while harnessing their individual strengths, such as easy manipulation and long coherence times. Recent experiments have shown the possibility of coherent energy exchange \cite{zhu_coherent_2011,kubo_storage_2012} on the single photon level and basic memory operations have also been demonstrated already in this context \cite{Wu_storage_2010,zhu_coherent_2011,kubo_storage_2012}.

\begin{figure}[!ht]
\centering
\includegraphics[width=0.45\textwidth]{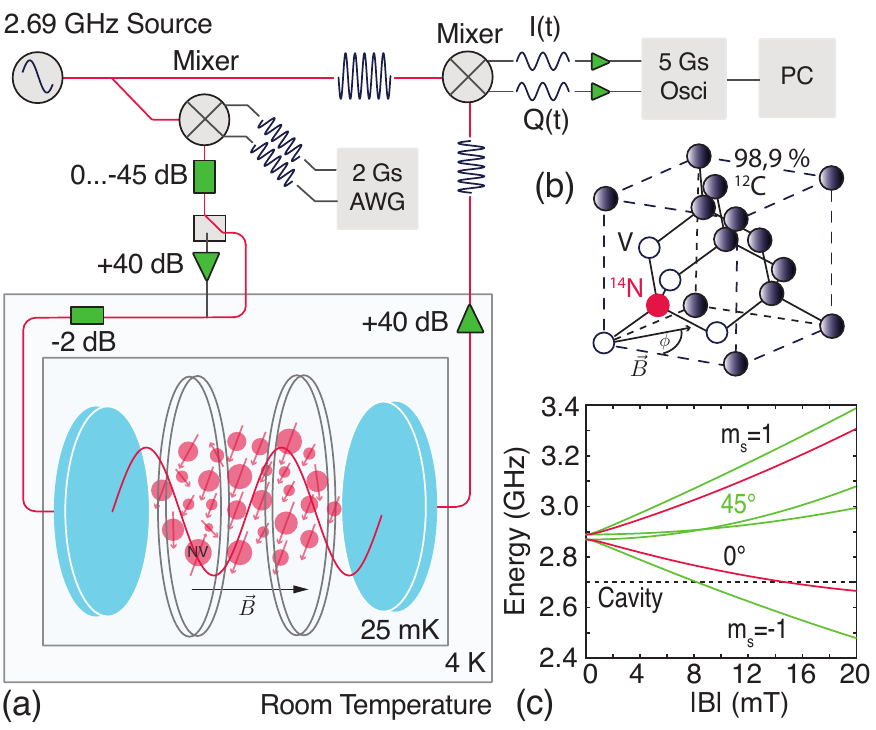}
\linespread{1.0}
\caption{\textbf{Experimental setup.} \textbf{(a)} A superconducting cavity loaded with an enhanced synthetic diamond crystal and surrounded by a three-dimensional d.c.~Helmholtz coil cage is placed in a dilution refrigerator operating at temperatures below 25~mK. Further details on the measurement scheme can be found in the appendices. \textbf{(b)} Structure of the nitrogen vacancy (NV) defect center ($S=1$) in diamond consisting of substitutional nitrogen atom and an adjacent lattice vacancy. \textbf{(c)} Zeeman tuning of the NV central spin transition frequencies for two fixed magnetic field angles $\phi$.}
\label{fig:figure1}
\end{figure}

The most pressing challenge in these hybrid systems remains to suppress spin dephasing induced by inhomogeneous spectral line broadening caused by the spin host material or by spin diffusion due to spin-spin interactions. The realisation of true multi-mode memories is however only possible in the presence of inhomogeneous spectral spin broadening such that their short memory times have to be actively recovered by echo refocusing techniques \cite{tyryshkin_electron_2012} or improved by the cavity protection effect \cite{putz_protecting_2014}.~Here we will present an alternative approach that circumvents the necessity for such recovery protocols, by exploiting the long-lived coherence of collective dark states \cite{zhu_observation_2014}.~Instead of employing~many~individual~broad~subensembles \cite{zhang_magnon_2015} we demonstrate here that such long-lived states can be engineered within the spin distribution by burning judiciously placed spectral holes, which creates isolated dark states protected against dissipation to~the~cavity~and~dephasing~in~the~remaining~bath~of~subradiant~modes.
\newpage
Our experimental system is composed of a superconducting resonator with a diamond crystal containing an ensemble of negatively charged nitrogen vacancy (NV) centre \cite{jelezko_observation_2004} electron spins magnetically coupled to it. The loaded cavity is then placed in a dilution refrigerator operating at low temperatures ($\leq$25~mK) see Fig.~\ref{fig:figure1}(a). The resonator is characterised at zero external magnetic field by transmission spectroscopy measurements and has a cavity linewidth $\kappa/2\pi=440\pm10$~kHz (HWHM) with a fundamental resonance at $\omega_c/2\pi$=2.691~GHz and a quality factor of $Q=3130$ (see Fig.~\ref{fig:rabi_split}). The diamond crystal has an approximate NV concentration of $\approx4 \times 10^{17}~$cm$^{-3}$ and is almost entirely polarised ($\geq$99 \%) at our refrigerator base temperature. The macroscopic spin ensemble in the cavity mode volume consists of $N\approx10^{12}$ NV spins. These effective two-level systems are then Zeeman shifted into resonance with the cavity by applying an external d.c. magnetic field tilted by 45$^\circ$ in the (100) plane of the crystal, at which two NV subensembles are brought into resonance with the resonator as shown in Fig.~\ref{fig:figure1}(c).

We observe a clear normal mode splitting and Rabi oscillations with a frequency $\Omega_R/2\pi=21.3\pm0.1$~MHz as well as a linewidth and decay rate $\Gamma/2\pi=2.9\pm0.1$~MHz (FWHM), by carefully probing the system with low intensities of $<$10$^{-6} $ photons per spin in the cavity (see Fig.~\ref{fig:rabi}).~Although the single spin-cavity coupling strength, $g_j$, is rather small ($\lesssim$10~Hz \cite{amsuss_cavity_2011}), the large number $N$ of weakly dipole-dipole interacting spins allows us to deeply advance into the strong-coupling regime ($\Omega_R\gg \Gamma\gg\kappa$) with a cooperativity $C\approx 20$. Such an ensemble of individual two-level systems coupled to a single mode cavity is described by the Tavis-Cummings model \cite{tavis_exact_1968}, which in the rotating wave approximation can be written as 
\begin{eqnarray}
\mathcal{H}=\hbar\omega_ca^{\dagger}a+\frac{\hbar}{2}\sum_{j=1}^N\omega_j \sigma_j^z+\hbar \sum_{j=1}^N g_j\left[\sigma^-_{j}a^{\dagger}+\sigma^+_j a\right]\!\!,\;
\label{eq:Hamilt_fun}
\end{eqnarray}
with bosonic creation (annihilation) operators $a^{\dagger}$~($a$) standing for the cavity mode whose frequency is $\omega_c$. The Pauli spin operators $\sigma^{\pm,z}_j$ are associated with the $j^{\text{th}}$ spin whose frequency is $\omega_j$. In such an ensemble of $N$ spins, sharing a single excitation, one finds a superradiant state \cite{dicke_coherence_1954} $\ket{B}=J^+\ket{G}$ and $N-1$ subradiant states $\ket{S}$ with collective spin operators $J^{\pm}=\sum_j^N \frac{g_j}{\sqrt{\sum_i^Ng_i^2}}\sigma_j^{\pm}$ and spin ground state $\ket{G}$.

\begin{figure*}[!ht]
\centering
\includegraphics[width=0.95\textwidth]{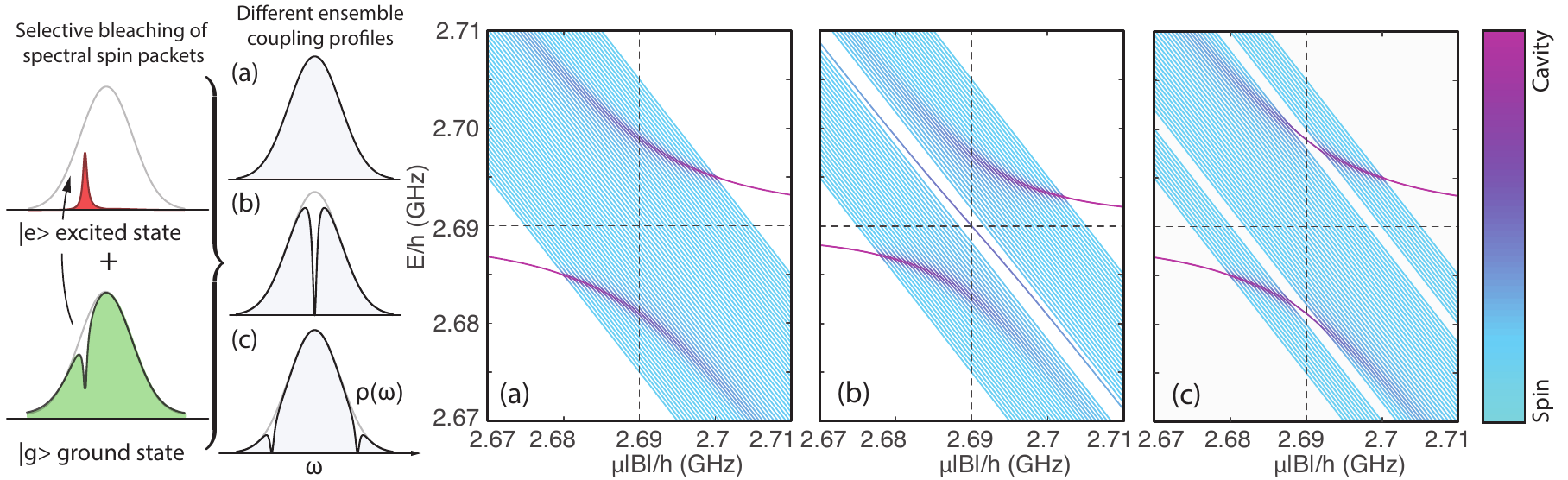}
\linespread{1.0}
\caption{\textbf{Visualisation of engineered collective dark states.} 
Spectral hole burning in narrow frequency windows bleaches spin components by bringing them into a mixture between their ground and excited state. The original smooth spin density $\rho(\omega)$ shown in \textbf{(a)} is then modified as is depicted in \textbf{(b-c)} \textit{(left panel)}. The calculated eigenenergy spectra of a cavity strongly coupled to a broadened spin ensemble is shown with cavity and spin contributions being indicated by the color gradient \textit{(colour bar on the right)}: two prominent polariton modes in a continuous bath of subradiant states are visible in the absence of hole burning \textbf{(a)}. Spectral hole burning at $\omega_s$ \textbf{(b)} and at $\omega_s\pm\Omega_R/2$ \textbf{(c)} creates long-lived resonant dark states. At these positions isolated dark state emerge which lie within the created holes and decouple from the remaining bath of subradiant states.}
\label{fig:figure3}
\end{figure*} 

The coupling of the cavity and spin ensemble gives rise to two polariton modes $\ket{\pm}=(\ket{1}_c\ket{G}_s\pm\ket{0}_c\ket{B}_s)/\sqrt{2}$ which are maximally entangled states between the cavity and the superradiant spin state. The hybridisation of the polariton modes or vacuum Rabi splitting \cite{thompson_observation_1992} is due to a collectively enhanced interaction $\Omega_R/2\approx\Omega=\sqrt{\sum_j^Ng_j^2}$ scaling approximately as $\sqrt{N}$, whereas subradiant states remain uncoupled and degenerate in the absence of spin broadening. In this sense an excitation that is stored in the subradiant space \cite{beige_driving_2000,Wu_storage_2010} will be protected from dissipation in the cavity mode and can be recovered by a spin echo, till eventually it decays with the single spin dissipation rate $\gamma$ \cite{tyryshkin_electron_2012}. The polariton modes and subradiant states, however, will not be entirely decoupled from each other in the presence of inhomogeneous spin broadening, corresponding to a variation of $\omega_j$ centered around a central spin frequency $\omega_s$. In our experiment the broadened spin ensemble has a linewidth $\gamma_{\text{inh}}/2\pi=9.4$~MHz~(FWHM) and a $q$-Gaussian spectral line shape $\rho(\omega)$  \cite{sandner_strong_2012}. This line broadening is a source of decoherence, which accelerates the evolution of an excitation stored in the superradiant spin state~into~the~bath~of subradiant~states.

\begin{figure*}[!ht]
\centering
\includegraphics[width=.9\textwidth]{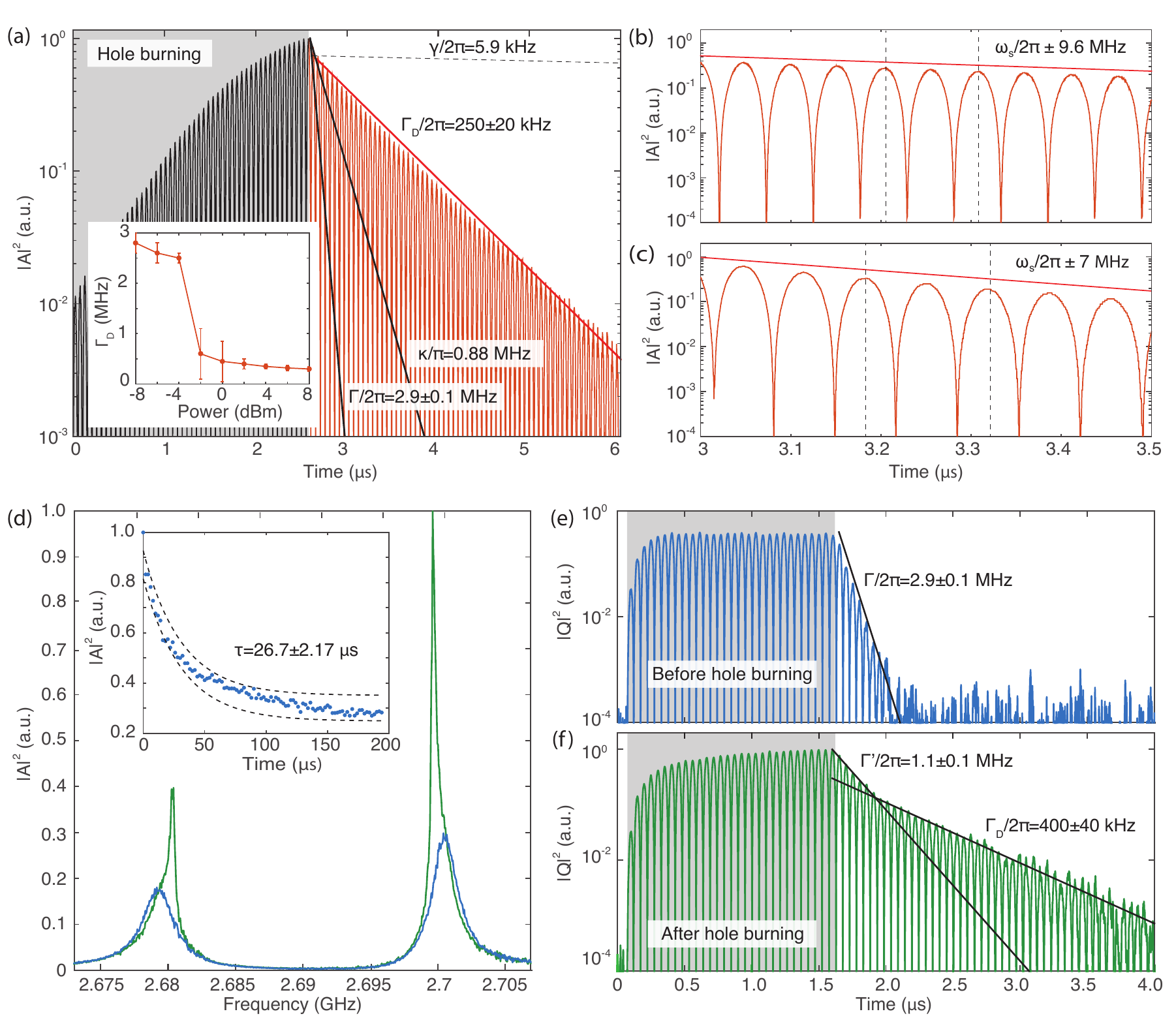}
\linespread{1.0}
\caption{\textbf{Spectral hole burning, dark state spectroscopy and dark state dynamics.} \textbf{(a)} A sinusoidal  pulse with modulation frequency, $2\pi\,9.6$~MHz $\approx \Omega_R/2$, and carrier frequency, $\omega_p=\omega_c=\omega_s$ \textit{(gray area)}, burns two spectral holes at $\omega_p/2\pi\pm9.6$~MHz of width $\Delta/2\pi=470$~kHz (FWHM). After the pulse intensity reaches a critical threshold \textit{(inset)} the decay rate is slowed down after the drive is switched off \textit{(red)}. For comparison we also plot the decay rates $\kappa$, $\Gamma$ and $\gamma$. \textbf{(b)} Close up of the damped Rabi oscillations within the time interval of $0.5~\upmu$s shown in \textbf{(a)}. The Rabi frequency can be controlled by varying the modulation frequency as is shown in \textbf{(c)}. \textbf{(d)} We scan the spectral transmitted steady state intensity through the cavity with low probe powers of $\approx 10^{-5}$ photons per spin: before \textit{(blue)} and $>$5$~\upmu$s after \textit{(green)} a hole burning pulse has decayed. Spectral holes were burned at $\approx\omega_s\pm\Omega_R/2$ and as a result two narrow peaks emerge directly on top of the polaritons due to the created dark states. The unequal amplitudes are attributed to an additional Fano resonance, although spins and cavity are in resonance $\omega_c=\omega_s$. The lifetime of the holes is measured by repeatedly probing the system with low intensities and monitoring the decaying amplitude \textit{(inset)}. \textbf{(e)} Linear dynamical response for a sinusoidally modulated weak pulse with a carrier frequency $\omega_p=\omega_c=\omega_s$ and a modulation frequency $\Omega_R/2$ \textit{(gray area)}. The substantially longer coherence time is proved by applying the same pulse as in \textbf{(e)}, $5~\upmu$s after the holes were burnt and the cavity was emptied. The errors in \textbf{(a-f)} correspond to the min and max values of the estimated decay rates and in the inset of \textbf{(d)} to the $1\sigma$ deviation.}
\label{fig:figure4}
\end{figure*} 

A solution to extend the coherence times and to circumvent spin dephasing has been pointed out in recent works \cite{diniz_strongly_2011,kurucz_spectroscopic_2011,krimer_non-markovian_2014}. The polariton modes $\ket{\pm}$ can be energetically decoupled from the bath of subradiant states by the ``cavity protection effect", and the total decay rate $\Gamma$ is substantially reduced in the limit of large coupling strengths $\Omega$. An implementation of this approach requires, however, extremely large coupling strengths $\Omega$ \cite{putz_protecting_2014} and reaches its fundamental limit $\Gamma=(\kappa+\gamma)/2$ only for $\Omega \rightarrow \infty$. At first sight it seems utterly impossible to go below this limit, since the decay rate of an excitation that is equally shared between the cavity and the spin ensemble is naturally bounded from below by the average of both decay rates. In the case at hand where the spin decay is much smaller than that of the cavity, $\gamma \ll \kappa$, a possibility to break this limit of $\Gamma=(\kappa+\gamma)/2$ would be to engineer polariton states that live mostly in the spins and only very little in the cavity. As theoretical predictions suggested recently \cite{krimer_hybrid_2015}, such ``collective dark states" can be created simply by spectral hole burning - a recipe which is conveniently implemented by bleaching or shelving away distinct spectral spin components of $\rho(\omega)$ through an externally applied pulse. As we will show below such a preparatory step is ideally suited as an all-purpose solution for reducing the decoherence which can in principle be followed by an arbitrary protocol that fits within the long lifetime of the spectral holes.


As we will see later on the optimal locations for burning narrow spectral holes of width $\Delta$ are at the centre $\omega_s$ of the spin ensemble and at the position of the polariton modes $\omega_s\pm\Omega_R/2$, as shown in Fig.~\ref{fig:figure3}(b) and (c), respectively. At these positions, states isolated from the bath of subradiant states emerge and lie within the spectral holes $\Delta$. In other words, the eigenstates at $\omega_s$ and $\omega_s\pm\Omega_R/2$ are an antisymmetric superposition of spins blue and red detuned with respect to the spectral hole. This fact distinguishes these dark states from the remaining subradiant states and we can write them as an antisymmetric state $\ket{A}\approx(\ket{\downarrow_B\uparrow_R}-\ket{\uparrow_B\downarrow_R})/\sqrt{2}$ coupled to the cavity mode 
\begin{eqnarray*}
 \ket{D}\approx\frac{1}{\sqrt{\Delta^2+2g_\mu^2}}\left( g_\mu\sqrt{2}\ket{A}\ket{0}_c+\Delta\ket{\downarrow_B\downarrow_R}\ket{1}_c \right)
\label{Hamilt_fun}
\end{eqnarray*}
with an effective cavity spin coupling strength $g_\mu$. On the other hand saturated spins in the spectral hole remain uncoupled and do not contribute. Such engineered dark states $ \ket{D}$ result in narrow peaks in the transmission, with a linewidth $\Gamma_D\geq\gamma$ directly related to the width of the spectral hole $\Delta$, which can be substantially narrower than cavity linewidth $\kappa$ as we show below.

In the following we engineer dark states at $\approx\omega_s\pm\Omega_R/2$ as in Fig.~\ref{fig:figure3}(c) and thereby introduce coherent long-lived Rabi oscillations between the spin ensemble and the cavity. We experimentally implement a high intensity hole burning pulse and study the reduced cavity decay of the hybridised system. The probe tone $\omega_p=\omega_c=\omega_s$ is modulated by a sinusoidal signal $\sin(\Omega_R\,t/2) \text{e}^{-i\omega_pt}$ and a Gaussian envelope resulting in two frequency components at $\omega_p\pm\Omega_R/2$ with $\Delta/2\pi=470$~kHz bandwidth. We use power values of up to 20 milliwatt corresponding to a steady state of $\approx10^{4}$ photons per spin in the cavity. This intensity is strong enough to bleach spin components selectively at the frequencies of the modulated drive signal by bringing them into a mixture of their ground and excited state and cancelling their effective spin-cavity interaction. Note that saturated spins will decay then slowly towards their ground state on a time scale of $\ge$10~ms, given by their spin lifetime \cite{amsuss_cavity_2011} of $T_1$=45~s which is shortened due to the Purcell effect \cite{e._m._purcell_proceedings_1946}. When increasing the hole burning pulse intensity as shown in the inset of Fig.~\ref{fig:figure4}(a) we find that dark states emerge above a certain power threshold. After we switch off the hole burning pulse shown in Fig.~\ref{fig:figure4}(a), we observe coherent Rabi oscillations with a transmitted intensity $|A(t)|^2$ through the cavity that decays substantially slower than for drive powers below the threshold (see Fig.~\ref{fig:figure4}(b) for more detail). In Fig.~\ref{fig:figure4}(c) we demonstrate that the hole burning procedure not only suppresses the decoherence but also allows us to control the Rabi flopping frequency when varying the position of the spectral holes.

In our system the best achievable decay rate of the engineered collective dark states is $\Gamma_D/2\pi=250\pm10$~kHz by creating spectral holes at frequencies $\omega_p/2\pi\pm9.6$ MHz, which is significantly below the fundamental limit reachable by the ``cavity protection effect" \cite{diniz_strongly_2011,kurucz_spectroscopic_2011,putz_protecting_2014,krimer_non-markovian_2014}. To observe the effect of hole burning and created dark states directly, we compare spectroscopic transmission measurements without hole burning and 5~$\upmu$s after a hole burning pulse has been applied and the cavity field has decayed. The spectrally resolved transmitted steady state intensity $|A(\omega_p)|^2$ is shown in Fig.~\ref{fig:figure4}(d) by scanning $\omega_p$ before and after spectral holes were burnt at positions equal to $\omega_s\pm\Omega_R/2$: as a result of the spectral holes two narrow dark states emerge directly on top of the polariton peaks. The created spectral holes and engineered dark states in Fig.~\ref{fig:figure4}(d) decay with a time constant $\tau=26.7\pm2.17$~$\upmu$s due to spin diffusion limiting the spectral hole lifetime. Note that in our experiment the spectral hole lifetime is more than a factor or four longer than the best achievable spin echo time $T_2=4.8\pm1.6$~$\upmu$s, although they are both limited by the same processes in our experiment (see appendices for further details).

To prove that this long-lived coherence can also be used for protocols that follow the hole burning procedure - such as for storing a low intensity excitation - we drive the system with a pulse $\sin(\Omega_R\, t/2) \text{e}^{-i\omega_pt}$ ($\omega_p=\omega_c=\omega_s$) corresponding to $<$$10^{-5}$ photons per spin in the cavity. Before the hole burning sequence we observe the unchanged system decay rate $\Gamma/2\pi=2.9\pm0.1$~MHz after switching off this weak drive - see Fig.~\ref{fig:figure4}(e). We then apply a hole burning pulse and 5~$\upmu$s after the signal has decayed the system is probed again. Remarkably, the Rabi oscillations need substantially longer time to set into the stationary state, which is a clear signature of the improved coherence time. After the pulse is switched off, two decays rates are discernible in the Rabi oscillations, see Fig.~\ref{fig:figure4}(f). We first observe a decreased decay rate, $\Gamma'/2\pi=1.1\pm0.1$~MHz, related to the time scale defined by the width of the polaritonic peaks which is, however, modified due to the hole burning and the interference between the energy stored in the spin ensemble and in the cavity. These dynamics are followed then by a crossover to a second even much slower asymptotic decay, $\Gamma_D/2\pi=0.4\pm0.04$~MHz, owing to the created dark states. These results demonstrate that our hole burning technique can be used as a convenient preparatory stage for many possible quantum information processing protocols. Since it is easy to implement and requires no active refocusing techniques, we expect this approach to find wide-spread use in hybrid quantum technology.

\begin{figure}[!ht]
\centering
\includegraphics[width=.45\textwidth]{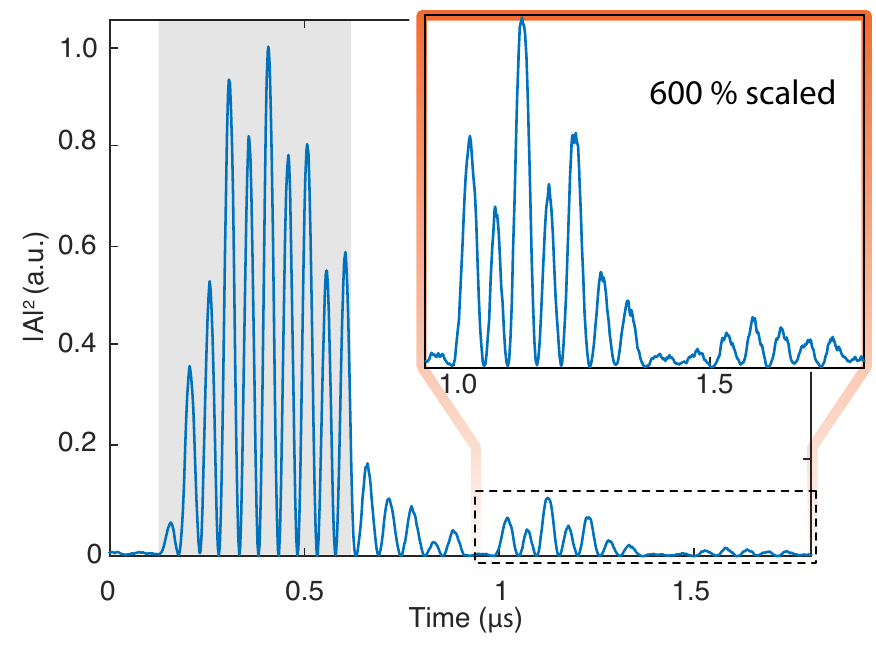}
\linespread{1.0}
\caption{\textbf{Engineering of multiple dark states.} We create four spectral holes and dark states at frequencies $\nu_1=\omega_s/2\pi\pm9$~MHz and $\nu_2=\omega_s/2\pi\pm10.8$~MHz close to the polariton modes. The response is probed with weak pulse intensities of $\approx10^{-5}$ photons per spin in the cavity. After the sinusoidally modulated drive with frequency $\Omega_R/2$ and carrier frequency $\omega_p=\omega_c=\omega_s$ \textit{(grey area)} is switched off, we observe a clear beating with $\Delta_{\nu_{21}}\approx~1.8$~MHz characterised by revivals of the damped Rabi oscillations.} 
\label{fig:figurecomb}
\end{figure}

To point a way towards one such future application, we also implemented multiple dark states by burning spectral holes with $\Delta/2\pi=300$~kHz bandwidth in the spin ensemble at $\nu_1=\omega_s/2\pi\pm9$ and $\nu_2=\omega_s/2\pi\pm10.8$~MHz whereby four isolated dark states in the proximity of the polariton modes are created. In Fig.~\ref{fig:figurecomb} we probe the dynamical response of this engineered system with low probe intensities corresponding to $<$$10^{-5}$ photons per spin in the cavity. We drive the cavity with a short sinusoidally modulated microwave pulse and observe a clear beating in the Rabi oscillations with beat frequency $\approx1.8$~MHz corresponding to the difference of both spectral hole frequencies $\Delta_{\nu_{21}}=1.8$~MHz. The beating produces two revivals in the Rabi oscillations which is a clear signature of the coherent nature of the created multiple dark states beating against each other. We have thus realized a key step towards constructing an on-chip all solid-state microwave frequency comb with long coherence times that may be a key component in future quantum memories and transducer networks.

\textbf{~\\Acknowledgements}\\
We would like to thank Arzhang Ardavan, Benedikt Hartl, Gerhard Kirchmair and Michael Trupke for helpful discussions. The experimental effort has been supported by the TOP grant of TU Wien. S.P. and A.A. acknowledge support by the Austrian Science Fund (FWF) in the framework of the Doctoral School ``Building Solids for Function" (Project W1243). D.O.K. and S.R. acknowledge funding by the FWF through Project No. F49-P10 (SFB NextLite).

\appendix
\renewcommand\thefigure{A\arabic{figure}}
\setcounter{figure}{0}    
\section{\Large{Appendices}}
\label{par:methods}
\subsection{\large{Experimental implementation}}
\textbf{The microwave cavity}
is loaded by placing the diamond sample on top of the $\lambda/2$ transmission line resonator. The super conducting microwave cavity is fabricated by optical lithography and reactive ion etching of a 200~nm thick niobium film sputtered on a 330 $\mathrm{\upmu m}$ thick sapphire substrate. The loaded chip is hosted and bonded to printed circuit board enclosed in a copper sarcophagus and connected to microwave transmission line.
\paragraph*{~\\}
\textbf{The spin ensemble is realised}
by enhancing a  tybe Ib high pressure high temperature diamond (HPHT) crystal containing an initial concentration of 200 ppm nitrogen with a natural abundance of $^{13}$C nuclear isotopes. We achieve a total density of $\approx6$ ppm NV centers by 50 hours of neutron irradiation with a fluence of 5$\times10^{17}$ cm$^{-2}$ and annealing the crystal for three hours at 900$^{\circ}$C. Excess nitrogen P1 centers ($S=1/2$) and additional lattice stress serve as main source of decoherence and spectral line broadening, by far exceeding the de-phasing due to the natural abundant 1.1$\%$ $^{13}$C spin bath. The diamond was initially characterised at room temperature by an optical laser scanning microscope.

\paragraph*{~\\}
\textbf{The negatively charged nitrogen vacancy} (NV) center is a paramagnetic impurity with electron spin $S=1$ which consists of a substitutional nitrogen atom and an adjacent vacancy in the diamond lattice. The electron spin triplet can be described by the Hamiltonian, $\mathcal{H}/h =DS_z^2+\mu \bm{B} \bm{S}$, with $\mu=28$ MHz/mT and a large Zero-field splitting $D= 2.877$ GHz corresponding to $hD/k_B\approx138$ mK allowing to thermally polarize the NV spin at finite temperatures of 25 mK up to 99\%. Due to the diamond lattice structure four different orientations of the adjacent vacancy are possible resulting in four NV subensembles. Magnetic field strengths of the order of $|B|=8$~mT are sufficient to bring cavity and spins in resonance, applied in the (100) crystallographic plane and in plane with the resonator. In the experiments presented in the main text the magnetic field is rotated by 45$^{\circ}$ in plane, at which only two subensembles are degenerate and in resonance with the cavity.
\paragraph*{~\\}
\textbf{The measurement scheme}
is an autodyne detection scheme for spectral hole burning and for measuring the transmitted intensity $|A(t)|^2$ through the cavity. The signal of a microwave source is split into two paths one serving as cavity probe tone and one as a local oscillator both with frequency $\omega_p$. The cavity probe tone is modulated by a frequency mixer and an arbitrary waveform generator (AWG) with 2 GS/s sampling frequency. The pulsed microwave probe tone can be attenuated up to -45 dB and routed by a fast switch through a high power amplifier with +40 dB gain. The microwave drive is then fed into the cryostat and attenuated by -2 dB on the 4 K stage allowing the application of up to 500 milliwatt power at the cavity input. The transmitted signal is fed into a low noise amplifier with +40 dB gain on the 4 K stage and mixed with the reference signal and both quadrature signals are recorded by an oscilloscope with 5 GS/s sampling frequency. From the measured quadrature's $I(t)$ and $Q(t)$ the transmitted microwave intensity $|A(t)|^2$ is calculated and plotted in Figs~3,4 and S2. The transmitted intensity through the cavity results in a steady state signal of $|A|^2=5 \times 10^{-4} \pm 2.5\times 10^{-7}$~(V$^2$) for a single shot of the 100 times averaged measurements shown in Fig.~3(e),(f) (where only one quadrature $|Q|^2$ is plotted) and Fig.~4.

\subsection{\large{Spectroscopic measurements}}
\textbf{Spin echo spectroscopy measurements} are employed to quantify the spectral hole lifetime. A Car-Purcell-Meiboom-Gill (CPMG) sequence \cite{carr_effects_1954} is employed to estimate the spin-spin relaxation time ($T_2$), and stimulated echo spectroscopy \cite{hahn_spin_1950} techniques are used to measure the spin-lattice relaxation time in the rotating frame ($T_{1\rho}$). The best achievable echo times in our experiment are $T_2=4.8\pm1.6$~$\upmu$s and $T_{1\rho}=6.4 \pm0.59$~$\upmu$s measured by CPMG and stimulated echos, respectively. We therefore conclude that the spin dissipation rate $\gamma=1/\tau=2\pi\,5.9$~kHz is dominated by spin diffusion in our experiment since $T_2\approx T_{1\rho}$. Although limited by the same process the spectral hole lifetime is more than a factor or four longer than $T_2$ and $T_{1\rho}$. This can be explained by the misalignment of the external d.c. magnetic field \citep{stanwix_coherence_2010} with respect to the NV axis and a bath of excess electron and nuclear spins in the host material.

\begin{figure}[!ht]
\centering
\includegraphics[width=.45\textwidth]{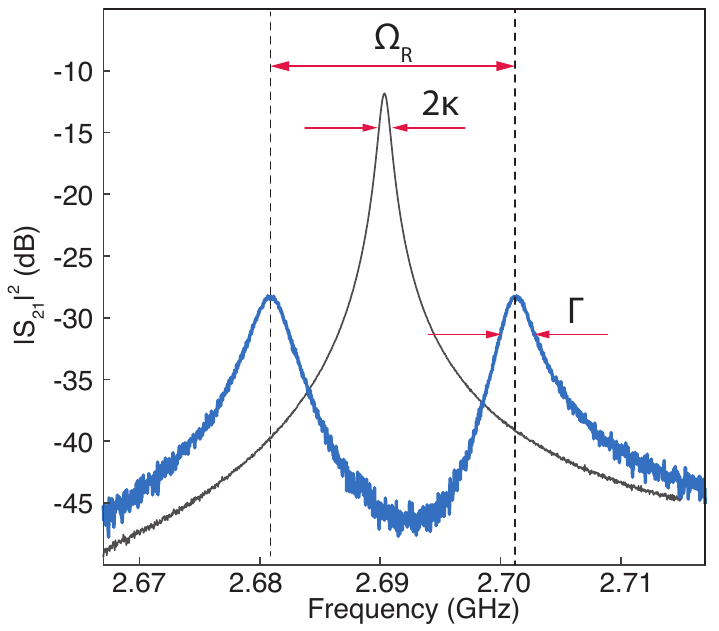}
\linespread{1.0}
\caption{\textbf{Transmission spectroscopy measurements.} The probe frequency $\omega_p$ is scanned and the cavity scattering parameter $|S_{21}|^2$ is measured by a vector network analyser. At zero external magnetic field the \textit{(black)} the cavity is largely detuned from the NV spin ensemble ($\omega_s\neq\omega_c$) and the bare cavity with a linewidth $\kappa/2\pi=440\pm10$~kHz (HWHM) with a fundamental resonance at $\omega_c/2\pi$=2.691~GHz and a quality factor of $Q=3,130$ is observed. A d.c. magnetic field of $|B|\approx 8$~mT is applied and a normal mode splitting \textit{(blue)} $\Omega_R/2\pi\geq19\pm0.1$~MHz is observed with a linewidth $\Gamma/2\pi=2.9\pm0.1$~MHz (FWHM) when the spin ensemble is in resonance with the cavity ($\omega_s=\omega_c$).}
\label{fig:rabi_split}
\end{figure}

\begin{figure}[!ht]
\centering
\includegraphics[width=.45\textwidth]{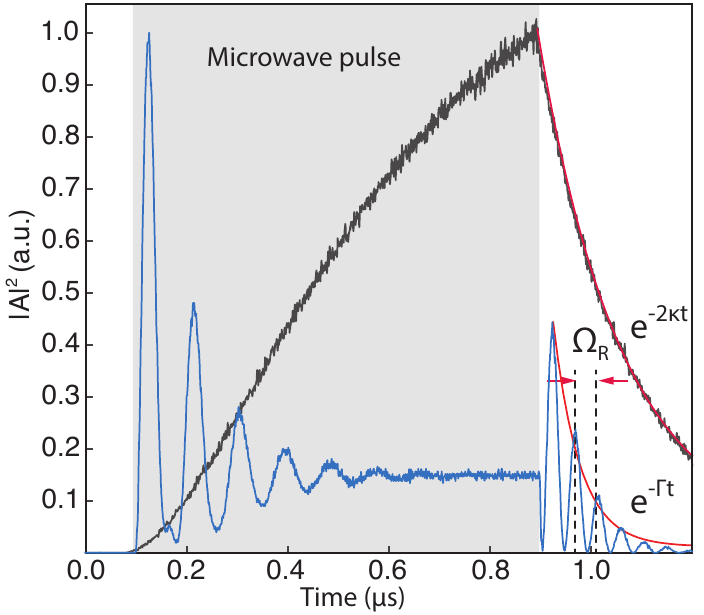}
\linespread{1.0}
\caption{\textbf{Time dependent cavity transmission.} A rectangular 800 ns long microwave pulse \textit{(gray area)} with carrier frequency $\omega_p=\omega_c=\omega_s$ is applied. The transmitted intensity $|A|^2$ is measured by an auto-dyne measurement scheme and the down converted signal is displayed. At zero external magnetic field ($\omega_p=\omega_c\neq\omega_s$) the bare and resonant cavity transmission \textit{(black)} is observed. After the drive is switched off the intensity decays with $\kappa=440\pm10$ kHz. The NV ensemble is brought into resonance with the cavity ($\omega_p=\omega_c=\omega_s$) by a Zeeman shift and the system hybridises. We observe clear Rabi oscillations \textit{(blue)} in the transmission signal with $\Omega_R/2\pi \geq 19\pm0.1$~MHz and a decay rate $\Gamma/2\pi=2.9\pm0.1$~MHz. Both signals have been normalised with respect to their maximal transmitted intensity.}
\label{fig:rabi}
\end{figure}

\subsection{\large{Theoretical analysis}}
\textbf{The modelling of the dynamics}
\label{Section_Volt_ampl}
after the hole burning process is done by deriving the Heisenberg operator equations for the cavity and spin operators, $\dot a=\frac{\text{i}}{\hbar}[{\cal H},a]-\kappa a$, $\dot \sigma_j^-=\frac{\text{i}}{\hbar} [{\cal H},\sigma_j^-]-\gamma \sigma_j^-$, respectively, where ${\cal H}$ stands for the Tavis-Cummings Hamiltonian given by Eq.~(1) of the main article. Although the hole burning is a nonlinear process, our primary aim here is a theoretical model which is capable of capturing the linear non-Markovian dynamics in the limit of weak driving powers after the holes in the spin density have been burnt. This allows us to simplify the equations by setting $\langle \sigma_j^z \rangle \approx -1$ (Holstein-Primakoff-approximation \cite{primakoff_many-body_1939}) and we derive the following linear set of first-order ODEs with respect to the cavity and spin amplitudes
\begin{equation}
\centering
\begin{split}
\dot{A}(t)  &=  -\left[\kappa+i(\omega_c-\omega_p)\right]A(t) + \sum_j g_j  B_j(t)-\eta(t), \\
\dot{B}_j(t) &=  -\left[\gamma+i(\omega_j-\omega_p)\right] B_j(t) - g_j A(t),
\end{split}
\label{Eq_bk_Volt}
\end{equation}
where $A(t)= \langle a(t)\rangle$ and $B_j(t) = \langle\sigma_j^-(t)\rangle$ with $\eta(t)$ being a time dependent drive term with a carrier frequency $\omega_p$.

\begin{figure}[ht!]
\includegraphics[width=.45\textwidth]{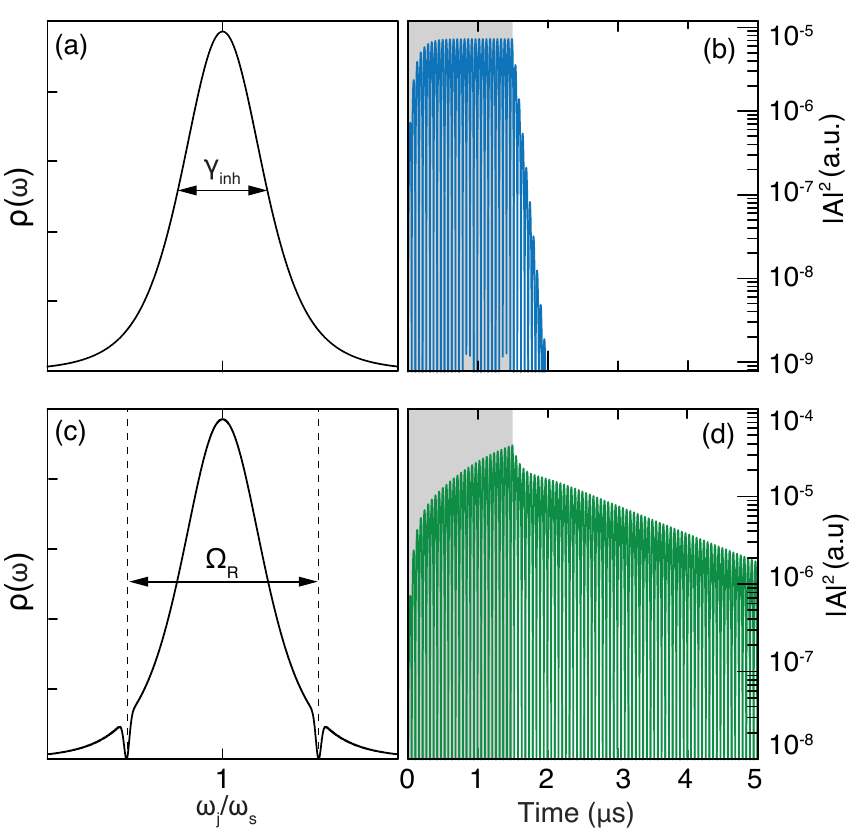}
\caption{\textit{(Left panel)}: spectral spin density modeled by a $q$-Gaussian distribution \cite{sandner_strong_2012,putz_protecting_2014,krimer_non-markovian_2014} without (a) and with two spectral holes (b) burned at $\omega_s\pm\Omega_R/2$. \textit{(Right panel)}: the corresponding dynamics under the action of a sinusoidal driving signal with small intensity, $\sin (\Omega_R\,t/2)\,\text{e}^{-\text{i} \omega_p t}$, with $\omega_p=\omega_s=\omega_c$. Gray (white) area indicates the time interval during which the driving signal is on (off).}
\label{fig_Gauss_0p7MHz_log_2holes_OmegaR_half}
\end{figure}

Owing to the large number of spins within the ensemble, we introduce a continuous spectral density as $\rho(\omega)=\sum_j^N g_j^2 \delta(\omega-\omega_j)/\Omega^2$, where $\Omega^2=\sum_j^Ng_j^2$ is the collective coupling strength of the spin ensemble to the cavity. Finally, we set up the so-called Volterra equation for the cavity amplitude, $A(t)=\int\limits_0^t d\tau {\cal K}(t-\tau) A(\tau)+{\cal F}(t)$, with the memory kernel function, ${\cal K}(t-\tau)=\int d\omega\rho(\omega) {\cal S}(\omega,t,\tau)$ (see \cite{putz_protecting_2014,krimer_non-markovian_2014} for details). The latter has a nontrivial structure and strongly depends on the exact shape of the spectral spin density $\rho(\omega)$. The system dynamics is then calculated by assuming a weak sinusoidal driving pulse, $\eta(t)\,$=$\,\sin (\Omega_R \, t/2)\,\text{e}^{-i \omega_p t}$, with the carrier frequency matching the resonance condition $\omega_p=\omega_s=\omega_c$, similarly to what is done in the experiment (see Figs.~3 (e),(f) in the main text). The resulting dynamics is displayed in Figs.~\ref{fig_Gauss_0p7MHz_log_2holes_OmegaR_half} (b),(d) for the case without and with hole burning. Note that we achieve good agreement with the experimental data.

\paragraph*{~\\}
\textbf{The connection}
between the physics of spectral hole burning and the collective dark states shown in Fig.~2 of the main article is made by solving the eigenvalue problem of our spin-cavity system and analyzing the resulting spectra. After substituting $A(t)=A \exp(\lambda t)$, $B_j(t)=B_j \exp(\lambda t)$ and $\eta=0$ into Eqs.~(\ref{Eq_bk_Volt}), we derive the complex eigenvalue problem for $\lambda$, which can be represented schematically as, ${\cal L} \psi=\lambda \psi$, with $\psi=(A, B_k)^T$. Note that in Fig.~2 of the main article, the value of $E=Re(\lambda)$ is depicted. Remarkably, here we take advantage of the previously established precise continuous form for the spin density $\rho(\omega)$ and discretise straightforwardly our problem by performing the following transformation, $g_\mu=\Omega \sqrt{\rho(\omega_\mu)/\sum_l\rho(\omega_l)}$. Since in total we deal with a sizeable number of spins ($N\approx 10^{12}$), we make our problem numerically tractable by dividing spins into many spin packets, so that $g_\mu$ represent a coupling strength within each spin packet rather than an individual spin coupling strength $g_j$. To provide an intuitive understanding how spectral hole burning reduces the decay rate we discuss here the case when spins at frequencies $\omega_s$ and $\omega_s\pm\Omega_R/2$ with respect to the central spin frequency $\omega_s$ in a frequency window of width $\Delta$ are saturated and remain uncoupled.

\label{biblio}
\bibliographystyle{mynaturemag}
\bibliography{ZotOutput.bbl}

\end{document}